\documentstyle[12pt]{article}

\newcommand{\equ}{\begin{equation} \bigskip}
\newcommand{\eequ}{\bigskip \end{equation}}
\vspace{3ex}

\begin{document}
\title{%
Analysis of the HELIOS-3 $\mu^+\mu^-$ Data within a Relativistic
Transport Approach%
\thanks{Work supported by BMBF and GSI Darmstadt.}}
\author{%
 W. Cassing$^1$, W. Ehehalt$^1$ and I. Kralik$^2$
\\$^1$ Institut f\"ur Theoretische Physik,
Universit\"at Giessen\\D-35392 Giessen, Germany \\
$^2$ Institute of Experimental Physics, Slovak Academy of Sciences \\
SR-04353 Kosice
}
\maketitle

\begin{abstract}
We present a nonperturbative dynamical study of $\mu^+\mu^-$ 
production in proton-nucleus and nucleus-nucleus collisions at SPS energies
on the basis of the covariant transport approach HSD. For p + W reactions at
200 GeV bombarding energy the
dimuon yield for invariant masses m $\leq $ 1.6 GeV is found to be
dominated by the decays of the $\eta, \rho, \omega$ and $\Phi$ mesons.
For 200 GeV/A S + W collisions, however, the dimuon yield shows an
additional large  contribution from $\pi^+\pi^-$, $K^+K^-$ and $\pi \rho$
channels. We find that for `free'
meson masses and form factors the experimental cross section is clearly
underestimated for S + W in the invariant mass range 
 0.35 GeV $\leq$ m  $\leq$ 0.65 GeV and that the HELIOS-3 data can only be
reproduced within a hadronic scenario, if the $\rho/\omega$-meson mass drops
with baryon density.
This finding suggests a partial restoration of chiral symmetry in S + W
collisions at SPS energies.

\end{abstract}

\vspace{1cm}


\newpage
The question of chiral symmetry restoration at high baryon density is 
of fundamental interest since a couple of years \cite{brownrho,chiral}, 
but a clear 
experimental evidence has not been achieved, so far. The enhancement of
strangeness production as e.g. seen in the AGS data for the $K^+/\pi^+$
ratio \cite{kplus} might be a signiture for such a transition \cite{Ehehalt},
however, other hadronic scenarios can be cooked up to describe this
phenomenon as well \cite{Braun}.  On the other hand,
dileptons are particularly well suited for an investigation of the
violent phases of a high-energy heavy-ion collision because they can leave the
reaction volume essentially undistorted by final-state interactions.
Indeed, dileptons from heavy-ion collisions  have
been observed by the DLS collaboration at the BEVALAC \cite{ro88,na89,ro89}
and by the CERES \cite{CERES} and HELIOS collaboration \cite{HELIOS,HELI2} 
at SPS energies. 

The enhancement of the low mass dimuon yield in S + W compared to p + W
collisions \cite{HELIOS} has been first
suggested by Koch et al. \cite{Koch} to be due to $\pi^+\pi^-$ annihilation.
Furthermore, Li et al. \cite{Li} have proposed
that the enhancement of the $e^+e^-$ yield in S + Au collision as observed
by the CERES collaboration \cite{CERES} should be
due to an enhanced $\rho$-meson production (via $\pi^+\pi^-$
annihilation) and a dropping $\rho$-mass in the medium. In fact, their
analysis - which was based on an
expanding fireball scenario in chemical equilibrium - could be confirmed
within the microscopic transport calculations in ref. \cite{Cass95};
however, also a more conventional approach including the increase of
the $\rho$-meson width in the medium due to the coupling of the
$\rho, \pi, \Delta$ and nucleon dynamics \cite{Herrmann,asakawa,Chanfray}
was found to be compatible with the CERES data.
In this paper, we will carry out a similar study as in ref. 
\cite{Cass95} but for the dimuon data of the HELIOS-3 collaboration
\cite{HELIOS} which provide independent information on dilepton production
with rather good statistics.

In continuation of the dilepton production studies 
in refs. \cite{Cass95,Wolf1,Wolf2,Wolf3} 
we describe the dynamics of proton-nucleus or nucleus-nucleus reactions
by a coupled set of covariant transport equations with scalar and
vector self-energies of all hadrons involved. Explicitly propagated are
nucleons, $\Delta$'s, N$^*$(1440), N$^*$(1535) resonances as well as
$\pi$'s, $\eta$'s, $\rho$'s, $\omega$'s, $\Phi$'s, kaons and K$^*$'s with
their isospin degrees of freedom. For more detailed information on the
self-energies employed we refer the reader to ref. \cite{Ehehalt}, where
the transport approach HSD\footnote{Hadron-String-Dynamics} is formulated
and applied to nucleus-nucleus collisions from SIS to SPS energies.

In this analysis we calculate $\mu^+\mu^-$ production taking into account
the contributions from nucleon-nucleon, pion-nucleon and pion-pion
bremsstrahlung, the Dalitz-decays 
$\eta \rightarrow \gamma \mu^+\mu^-$ 
and $\omega \rightarrow \pi^0 \mu^+\mu^-$, the
direct dimuon decays of the vector mesons $\rho, \omega, \Phi$ 
as well as $\pi^+\pi^-$, $K^+K^-$ and $\pi \rho$ production channels. 
The nucleon-nucleon,
pion-nucleon and $\pi \pi$ bremsstrahlung are evaluated in a phase-space
corrected soft photon approximation as 
described in refs. \cite{Wolf1,Wolf3}. Though this limit is questionable
for the bombarding energy of interest, the soft photon approximation
provides an upper limit for the bremsstrahlung contribution \cite{Eggers},
which will turn out to be small in comparison to the meson decay channels.

The pion annihilation  
proceeds through the $\rho$-meson which decays into a
virtual massive photon by vector meson dominance. 
The cross section is
parametrized as  \cite{Wolf1,Wolf3,GaleK87} 
\begin{equation}
\sigma^{\pi^+\pi^-\rightarrow \mu^+\mu^-}(m) = \frac{4\pi}{3}
  \left( \frac{\alpha}{m} \right)^2 \sqrt{1-\frac{4 m_\pi^2}{m^2}} 
 \  |F_\pi(m)|^2 \; ,
\label{piann}
\end{equation}
where the free form factor of the pion is approximated by
\begin{equation} 
|F_\pi(m)|^2 = \frac{m_\rho^4}{(m^2-m_\rho {}^2)^2+m_\rho^2 \Gamma_\rho^2} 
  \; .
\end{equation}
In eq. (2) $m$ is the dilepton invariant mass, $\alpha$ is the fine structure
constant, and
\[
m_\rho = 775 MeV \:,  \hspace{2em} \Gamma_\rho = 118 MeV \; .
\]
So far, eq.(\ref{piann}) describes the free pion annihilation cross section;
possible medium modifications of the $\rho$-meson will be discussed below.
 
The cross section for the $K^+K^-$ annihilation is
parametrized as  \cite{liko} 
\begin{equation}
\sigma^{K^+K^-\rightarrow \mu^+\mu^-}(m) = \frac{4\pi}{3}
  \left( \frac{\alpha}{m} \right)^2 \sqrt{1-\frac{4 m_K^2}{m^2}} 
 \  |F_K(m)|^2 \; ,
\label{kann}
\end{equation}
where the free form factor of the kaon is approximated by
\begin{equation} 
|F_K(m)|^2 = \frac{1}{9} \frac{m_\Phi^4}{(m^2-m_\Phi^2)^2+m_\Phi^2 \Gamma_\Phi^2} 
\end{equation}
with
\[
m_\Phi = 1020 MeV \:,  \hspace{2em} \Gamma_\Phi = 4.43 MeV \; .
\]

The cross section for dimuon production in $\pi^+\rho^-, \ \pi^-\rho^+$ 
scattering is given by the Breit-Wigner cross section \cite{Physrep}
for the formation of
a $\Phi$-meson times the partial decay of the $\Phi$ into a $\mu^+\mu -$
pair,
$$
\sigma^{\pi^+\rho^- \rightarrow \mu^+\mu^-}(m) = \frac{\pi}{k^2}
   \frac{\Gamma_t^2}{(m - m_\Phi)^2 + \Gamma_t^2/4} B_{in} B_{out} ,
$$
$$
\Gamma_t(k) = \Gamma_\Phi \left( \frac{k}{k_r} \right)^3 
\frac{k_r^2 + q^2}{k^2 + q^2}  ,
$$
$$
q^2 = (m_\Phi - m_\rho - m_\pi)^2 +  \Gamma_\Phi^2/4 , 
$$
$$
k_r^2 = \left(m_\Phi^2 - (m_\rho + m_\pi)^2 \right) \left( m_\Phi^2
- (m_\rho - m_\pi)^2 \right) /(4 m_\Phi^2) , 
$$
\begin{equation}
k^2 = \left(m^2 - (m_\rho + m_\pi)^2 \right) \left( m^2
- (m_\rho - m_\pi)^2 \right) /(4 m^2) , 
\end{equation}
with $B_{in}$ = 0.13 and $B_{out}$ = 2.5 $\times 10^{-4}$. We note that
the resonance approximation (5) also works very well for $K^+ K^-$ annihilation
as described by eq. (3) when including a further spin factor of 3 and $B_{in}$
= 0.49. 

The $\omega$  Dalitz-decay, furthermore,  is given by \cite{Landsberg}:
\begin{eqnarray}
\frac{d\Gamma_{\omega \rightarrow \pi^0 \mu^+\mu^-}}{dm}
= \frac{\alpha}{3\pi} \frac{\Gamma_{\omega \rightarrow \pi^0 \gamma}}{m}
\left(1 - \frac{4 m_\mu^2}{m^2} \right)^{1/2} \left( 1 + 2 \frac{m_\mu^2}{m^2}
\right )   \nonumber \\
\times \left( (1+\frac{m^2}{m_\omega^2 - m_\pi^2})^2 - 
\frac{4 m_\omega^2 m^2}{(m_\omega^2-m_\pi^2)^2}
 \right)^{3/2} |F_{\omega \rightarrow \pi^0 \mu^+\mu^-}(m)|^2 \; ,
\end{eqnarray}
where the form factor is parametrized as 
\begin{equation}
F_{\omega \rightarrow \pi^0 \mu^+\mu^-}(m) = \frac{\Lambda_s^4}{(\Lambda_s^2-
 m^2)^2
+ \Lambda_s^2 \Gamma_s^2}
\end{equation}
with
\begin{equation}                            \label{omeform}
 \Lambda_s= 0.60 \mbox{ GeV} \; , \ \Gamma_s = 75 \mbox{ MeV} .
\end{equation}  
The form (7) for $F_\omega$ is actually fitted to the dimuon data for
p + W at 200 GeV for 0.4 GeV $\leq$ m $\leq$ 0.6 GeV and 
slightly deviates from the parametrization in
\cite{Landsberg} (with $\Lambda_s = 0.65 \pm 0.03$ GeV) but is still 
compatible with the data presented in \cite{Landsberg}. We note that this
formfactor is of interest in its own since it shows a significant
enhancement compared to the vector-meson-dominance model \cite{Landsberg}.
More accurate experimental data on this formfactor as well as further
theoretical studies will be needed to clarify the underlying physics. 

The direct decays of the vector mesons to $\mu^+\mu^-$ are taken as 
\begin{equation}
\frac{d\sigma}{dm^2}(m) = \frac{1}{\pi} \frac{m_V \Gamma_V}{(m^2-m_V^2)^2
+ m_V^2 \Gamma_V^2} \frac{\Gamma_{V \rightarrow \mu^+\mu^-}}{\Gamma_{tot}}
\end{equation}
with $\Gamma_{\Phi \rightarrow \mu^+\mu^-}/\Gamma^{tot}_\Phi $ =
2.5$\times  10^{-4}$, $\Gamma_{\omega \rightarrow \mu^+\mu^-}/\Gamma^{tot}_\omega$
= 7.1 $\times 10^{-5}$ and $
\Gamma_{\rho \rightarrow \mu^+\mu^-}/\Gamma^{tot}_\rho$ = 4.6 $\times 10^{-5}$.
Though of minor importance we also include the direct decay of the $\eta$-meson
with $\Gamma_{\eta \rightarrow \mu^+\mu^-}/\Gamma^{tot}_\eta$
= 5.7 $\times 10^{-6}$. The charged particle multiplicity $N_c$ - which also 
provides the
normalization of the experimental data - is recorded in the pseudorapidity interval
3.7 $\leq \eta \leq$ 5.2.

Transport models similar to ours have been applied for $e^+e^-$ 
production at BEVALAC energies in refs. \cite{Ko,Xiong,Toneev,Tonev}
and at AGS energies in \cite{Frank}. The first full microscopic studies for
 $e^+e^-$
production at SPS energies have been reported in \cite{Cass95}; the same
covariant transport approach is also used in our present analysis.

We have calculated the $\mu^+\mu^-$ yield for p + W at 200 GeV using
the experimental cuts in rapidity y and transverse mass $m_T$ 
of the dimuon source,
\begin{equation}
m_T \geq 4 (7-2y), \hspace{1cm} m_T \geq \sqrt{(2 m_\mu)^2 + \left(
\frac{15 \  GeV/c}{\rm{cosh} (y)} \right)^2},
\end{equation}
as well as the experimental resolution in energy  \cite{HELIOS}.
  
Our results are displayed in Fig. 1 for p + W at 200 GeV in comparison with the
HELIOS-3 data \cite{HELIOS}. In the upper part of Fig. 1 we show the calculated
results with an energy resolution of $\Delta_M$ = 15 MeV, while the experimental
resolution was used in the lower part of Fig. 1 in order to demonstrate
the sensitivity of the spectra to the actual mass resolution in the
experimental data. The full
solid curves in Fig. 1 display the sum of all individual contributions which is
dominated by the decays  of the mesons. The bremsstrahlung contributions
($\pi $N, pN and $\pi \pi$ channels summed up in the single line denoted by
'brems') as well as $\pi^+\pi^-$ annihilation (short dashed line) 
are of minor importance for
the proton induced reaction in full analogy to the CERES $e^+e^-$ data
\cite{CERES,Cass95}. The very good reproduction of the HELIOS-3 data,
where  medium effects are negligible, allows us to perform a related analysis
of nucleus-nucleus collisions with particular emphasis on medium properties
of the reaction channels involved. 
 
We thus go over to the system S + W at 200 GeV/A where we expect to
describe the global reaction dynamics with the same quality as for S + Au
at 200 GeV/A (cf. Fig. 2 of \cite{Cass95}). The results of our 
calculation, where no medium effects are incorporated for all mesons, are 
displayed in Fig. 2 (thick solid line) in comparison to the data 
\cite{HELIOS}. The individual contributions from bremsstrahlung channels
(denoted by 'brems'),
$\eta$ and $\omega$ Dalitz decays as well as direct vector meson decays
are explicitely indicated in Fig. 2.
Concentrating on the invariant mass range m $\leq $ 1 GeV we find the
experimental yield to be underestimated for 0.35 GeV $\leq$ m $\leq$ 0.65 GeV
and slightly overestimated for m $\approx$ 0.8 GeV.
In the $\Phi$-mass region the cross section includes the direct $\Phi$
production channels as well as those from $\pi \rho$ and $K^+K^-$ 
collisions\footnote{The kaon production channels $\pi \pi \rightarrow K \bar{K}$
are included in the transport calculation}.
The  $\pi^+\pi^-$ annihilation component (short dashed line) 
is found to be of the same order of magnitude as the direct $\rho$-decay
(long dashed line) within the experimental cuts and significantly smaller 
than in case of the CERES data \cite{Cass95} - which focus on the mid-rapidity
regime - while the additional bremsstrahlung contributions again
are of minor importance.

Since the reproduction of the HELIOS-3 data in Fig. 2 is rather poor
we will now explore possible in-medium modifications of the mesons as
advocated in refs. \cite{Cass95,Wolf3,liko,Ko,Shakin}.
Before going over to the actual model,
we show in Fig. 3 for S + W at 200 GeV/A 
the $\rho$-mass distribution versus the baryon density
$\rho_B$ in units of $\rho_0 \approx 0.168 fm^{-3}$, where is baryon density
is computed in the local rest frame, i.e. $\rho_B^2(x) = j_\mu j^\mu$, while
$j_\mu(x)$ denotes the Lorentz invariant baryon current. In Fig. 3 
the solid line represents the
distribution of $\rho$'s produced in baryon-baryon and pion-baryon
collisions whereas the dashed line corresponds to those from $\pi^+\pi^-$
annihilation. We note that the distributions in Fig. 3 are recorded at the
creation time of the $\rho$-mesons, whereas the distribution at the actual
decay time is slightly shifted to smaller density. Furthermore, only those
$\rho$-mesons have been considered which fulfill the experimental cuts (10)
in order to provide a close connection to the HELIOS-3 experiment.
These distributions clearly indicate that most of the $\rho$'s are produced
at nonzero baryon density and even for $\rho_B \geq 3 \rho_0$.
 
We now turn to the possible modifications of the $\rho$-meson in the medium.
From QCD inspired models \cite{brownrho,Shakin} or estimates based on
QCD sum rules \cite{hatsuda,asaka} it has been predicted that
the $\rho$-meson mass decreases with density. In order to explore the 
compatibility of such scenarios with
the HELIOS-3 data,  we have performed calculations with a medium-dependent 
$\rho$- (and $\omega$)-mass according to Hatsuda and Lee \cite{hatsuda}, i.e.
\begin{equation}
m_\rho^* \approx m^0_\rho \ (1 - 0.18 \rho_B/\rho_0) 
\geq m_u + m_d \approx 14 MeV ,
\end{equation}
where $\rho_B(t)$ is the actual baryon density during the decay of the
$\rho$-meson. A similar scaling with density has also been predicted 
by Shakin et al. \cite{Shakin}. 
Since the width of the $\rho$-meson is determined by the decay into two pions,
which are propagated without selfenergies as in \cite{Ehehalt}, we have
parametrized the in-medium width $\Gamma^*_\rho$ as
\begin{equation}
\Gamma^*_\rho = \Gamma_\rho \ \frac{m^{*2}_\rho - 4 m_\pi^2}{m^2_\rho 
- 4 m_\pi^2} \ \Theta(m^*_\rho - 2 m_\pi) .
\end{equation}
The dropping of the
$\rho$-meson mass is associated with a scalar self-energy of the meson which
is determined by the local baryon density; the propagation of the
`quasi-particle' with effective mass m$^*_\rho$ thus couples to the baryon
current during the expansion, and the meson becomes `on-shell' asymptotically
due to a feedback of energy from the mean fields, which thus ensures
that the total energy of the system is conserved.

The results of this simulation are shown in Fig. 4 in comparison with
the HELIOS-3 data.
The $\pi^+\pi^-$ annihilation component (short dashed line) now shows a peak 
at invariant masses m $\approx$ 0.6 GeV  and leads to a significant enhancement 
of low mass dileptons. A further enhancement at lower masses is also
provided by the direct $\rho$-decays (long dashed line) 
due to their shifted masses. In
comparison to Fig. 2 the experimental dimuon spectrum is now much better
reproduced for invariant masses m $\leq$ 1 GeV; 
the missing yield for m $\geq$ 1.2 GeV might be due to direct charm production
channels or higher meson resonances which are not explicitly included in
the present calculations.

Of further interest is the experimental rapidity dependence of the
dimuon spectra. Here, explicit data are available within the cuts (10) for
y $\leq$ 3.9, 3.9 $\leq$ y $\leq$ 4.4, and y $\geq$ 4.4 . The comparison
of our calculations with the data is shown in Fig. 5 for the
rapidity cuts 3.9 $\leq$ y $\leq $ 4.4 and y $\geq$ 4.4. The dotted 
lines correspond to calculations with
free meson masses and formfactors whereas the solid lines are the result
within the dropping meson mass scenario (11) with the in-medium width (12). 
Again the computations
with free meson masses fail to reproduce the data especially for the 
lower rapidity interval from 0.35 - 0.65 GeV
and about 0.8 GeV whereas the agreement is remarkably
fine when including the shift of the meson masses with density. 

As advocated by Haglin \cite{Haglin} the width of the $\rho$-meson should
again increase in a dense mesonic environment 
due to the mesonic interaction channels. In order to explore such effects
we have performed calculations with a dropping $\rho$-mass according to (11),
however, keeping its width at the free value, i.e. $\Gamma^*_\rho = \Gamma_\rho$.
The results of this 'theoretical experiment' are displayed in Fig. 5
in terms of the dashed lines. Whereas for y $\geq$ 4.4 the agreement with the
data improves, the $\mu^+ \mu^-$ yield is slightly underestimated for m
$\approx$ 0.5 GeV in the lower rapidity interval. We thus cannot extract a clear
evidence for a reduced width of the $\rho$-meson at high baryon density from
the present experimental data.

Within the same spirit we have also investigated, if the broadening of the
 $\rho$-meson
- without a shift of its pole - according to Herrmann et al. \cite{Herrmann},
which is due to the coupled $\pi$ and $\Delta$-hole dynamics, 
might improve the description of
the data. However, contrary to the $e^+e^-$ spectra of the CERES-collaboration
(cf. ref. \cite{Cass95}), there is no significant enhancement of the spectra
for 0.3 GeV $ \leq$ m $\leq$ 0.6 GeV (in comparison to Fig. 2) such that
this mechanism can no longer be considered as dominant. Furthermore,
since the pion densities in the rapidity regime of interest are large
compared to the baryon densities (cf. Fig. 2 of \cite{Cass95}), 
the pion selfenergy effects from $\Delta$-hole
loops are most likely overestimated \cite{Gerry}.

In summary, we have studied $\mu^+\mu^-$ production in proton and heavy-ion 
induced reactions at 200 GeV/A on the basis of the  
covariant transport approach HSD \cite{Ehehalt}.
We have incorporated
the contributions from proton-nucleon, pion-nucleon and pion-pion
bremsstrahlung, the Dalitz-decay of the
$\Delta$, $\eta$ and $\omega$ as well as $\pi^+\pi^-$ annihilation
and the direct dilepton decay of the vector mesons $\rho, \omega, \Phi$
as well as $K^+K^-$ and $\pi \rho$ channels.
It is found that for p + W at 200 GeV the mesonic
decays almost completely determine the dilepton yield, whereas in S + W
reactions the $\pi^+\pi^-, K^+ K^-$ annihilation channels and $\pi \rho$
collisions contribute substantially.
The experimental data taken by the HELIOS-3
collaboration \cite{HELIOS} are underestimated by the calculations for 
invariant masses 0.35 GeV $\leq$ m $\leq$ 0.65 GeV when
using free form factors for the pion and $\rho$-meson.
 
We have, furthermore, examined if a shift of the $\rho$- (and $\omega$)-mass 
according to QCD sum rules - as suggested by Hatsuda and Lee \cite{hatsuda}
or Shakin et al. \cite{Shakin} - is compatible with the data as in case of
the $e^+e^-$ spectra of the CERES collaboration \cite{Cass95}. Indeed the
description of the $\mu^+\mu^-$ spectrum improves substantially within this
'dropping mass' scenario at all rapidity bins. 
However, the effects at forward rapidities are
not as pronounced as close to midrapidity where the dilepton yield is
clearly dominated by the $\pi^+\pi^-$ annihilation channel (cf. Figs. 3-5 
of \cite{Cass95}). Nevertheless, the explicit rapidity dependence of the
experimental dimuon spectra, which is in line with the dropping meson 
mass scenario, provides a strong hint for the partial restoration of chiral 
symmetry at high baryon density.   
Data with high statistics especially at midrapidity or alternatively 
central collisions of Pb + Pb at
SPS-energies are expected to allow for more definite conclusions about the
chiral phase transition of hadronic matter at 4-6$\times \rho_0$ baryon
density.

\vspace{3ex}
The authors gratefully acknowledge many helpful discussions with 
E. L. Bratkovskaya, K. Haglin, C. M. Ko, U. Mosel, H. J. Specht, and
Gy. Wolf. They
are especially indepted to A. Drees for many hints and fruitful suggestions
throughout the course of this analysis.

\vspace{3cm}
\noindent
{\Large \bf Figure Captions}

\vspace{1.0cm}
\noindent
{\bf Fig. 1:} Comparison of our calculations for the differential dimuon
spectra (thick solid line) with the experimental data \cite{HELIOS} 
for p + W at 200 GeV. Upper part: with an energy resolution
of 15 MeV; lower part: with the experimental resolution. 
The individual contributions
from the vector mesons and Dalitz decays as well as bremsstrahlung
channels are directly indicated in the figure. 
The $\pi^+\pi^-$ annihilation 
channel and the decay from $\rho$-mesons produced in baryon-baryon and
meson-baryon collisions are shown by the short dashed and long dashed lines,
respectively.
 
\vspace{1.0cm}
\noindent
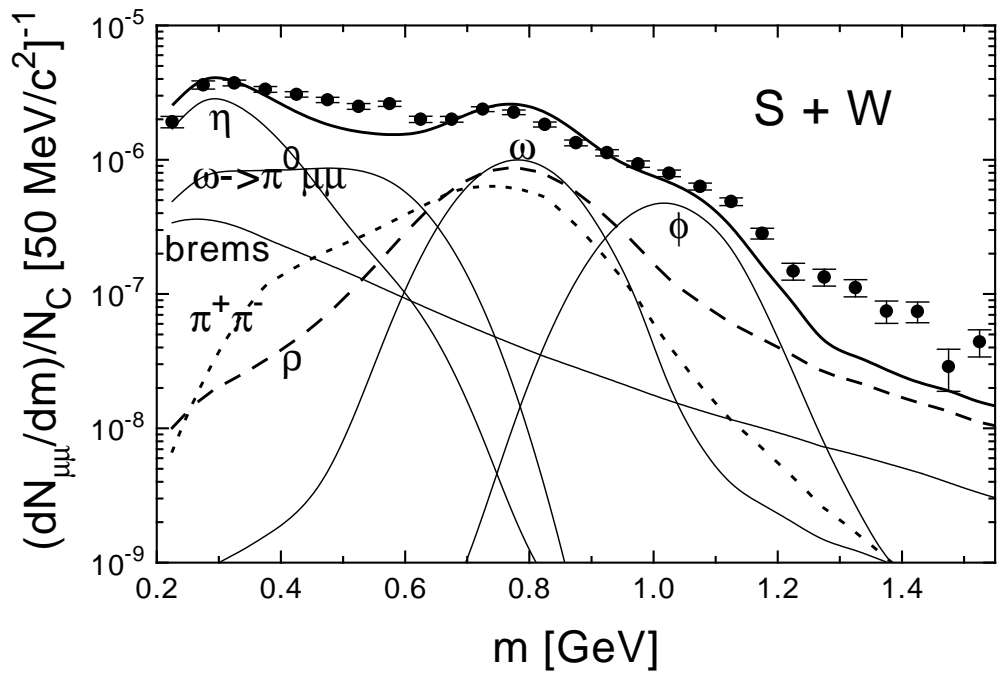
{\bf Fig. 2:} Comparison of our calculations for the differential dimuon
spectra (upper thick solid line) with the experimental data \cite{HELIOS} 
for S + W at 200 GeV/A when employing no medium modification
of the mesons. 
The $\pi^+\pi^-$ annihilation 
channel and the decay from $\rho$-mesons produced in baryon-baryon and
meson-baryon collisions are shown by the short dashed and long dashed lines,
respectively.

\vspace{1.0cm}
\noindent
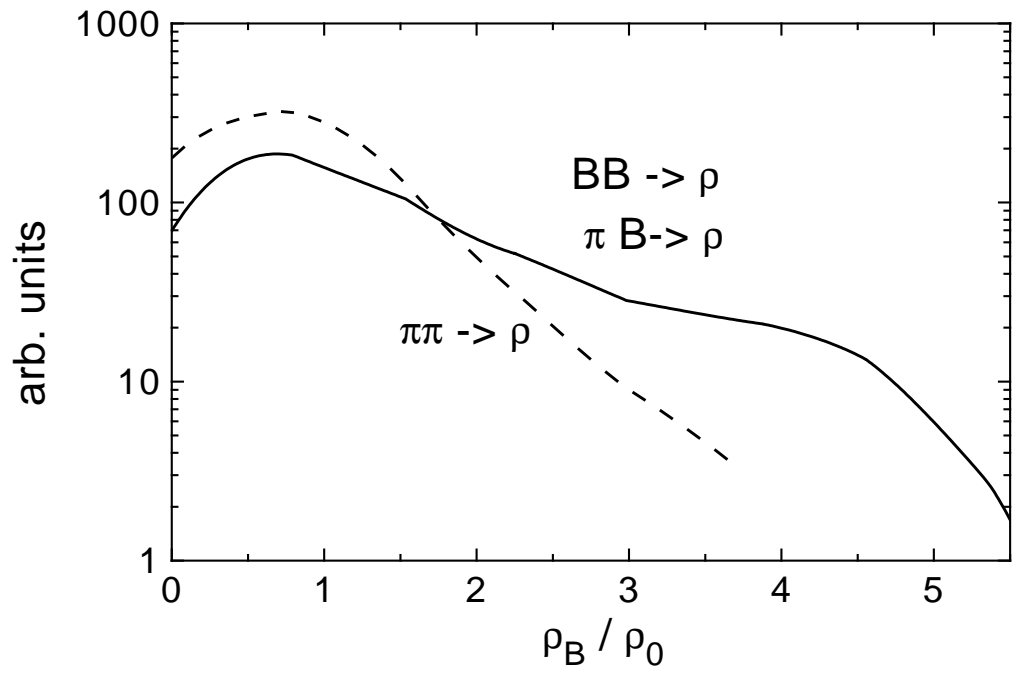
{\bf Fig. 3:} The $\rho$-mass distribution as a function of the baryon
density $\rho_B$ (in units of $\rho_0$) for a central S + W reaction at 200 GeV/A 
within the HELIOS-3
acceptance (10). Solid line: $\rho$-mesons from baryon-baryon and meson-baryon
reaction channels; dashed line: $\rho$-mesons from $\pi^+\pi^-$ annihilation.

\vspace{1.0cm}
\noindent
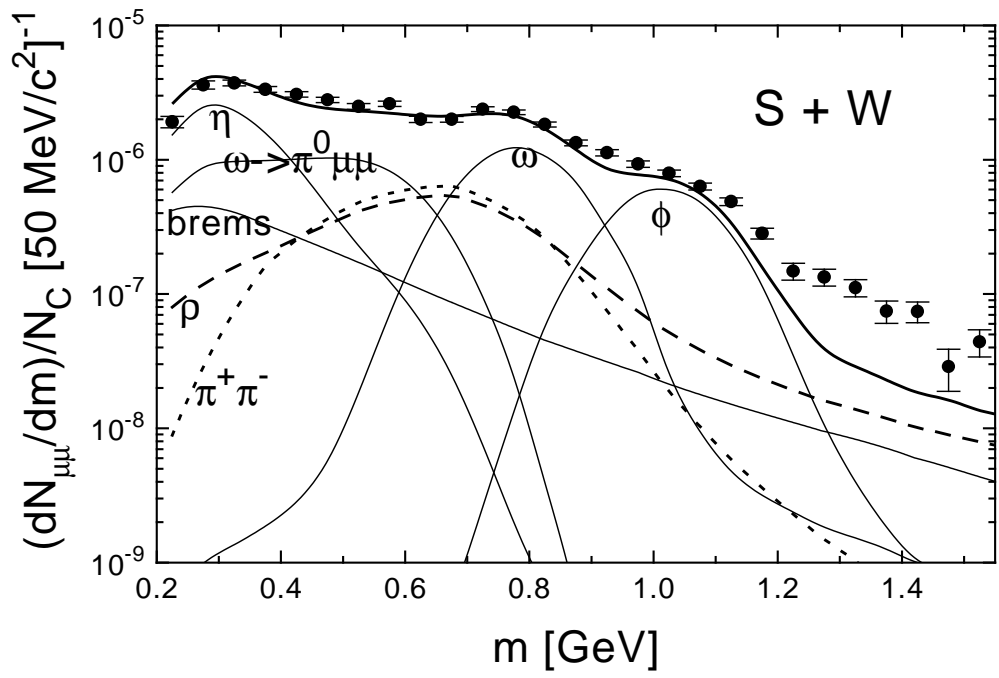
{\bf Fig. 4:} Comparison of our calculations for the differential dimuon
spectra (thick solid line) with the experimental data \cite{HELIOS} 
for S + W at 200 GeV/A when including a shift of the
$\rho$-meson mass (11) according to the prediction  by Hatsuda and Lee
\cite{hatsuda} and a decrease of its width according to (12).
The $\pi^+\pi^-$ annihilation 
channel and the decay from $\rho$-mesons produced in baryon-baryon and
meson-baryon collisions are shown by the short dashed and long dashed lines,
respectively.

\vspace{1.0cm}
\noindent
{\bf Fig. 5:} Result of our calculations for the differential dimuon
spectra  for the  rapidity cuts 3.9 $\leq$ y $\leq$ 4.4 and y $\geq$ 4.4 
for S + W at 200 GeV/A in comparison to the HELIOS-3 data \cite{HELIOS}.
The dotted lines correspond to a calculation with free meson masses and formfactors,
the solid lines to a calculation with a dropping $\rho$-mass according to (11) and
width (12), whereas the dashed line is obtained in the 
dropping $\rho$-mass scenario (11) with the free $\rho$-meson width.
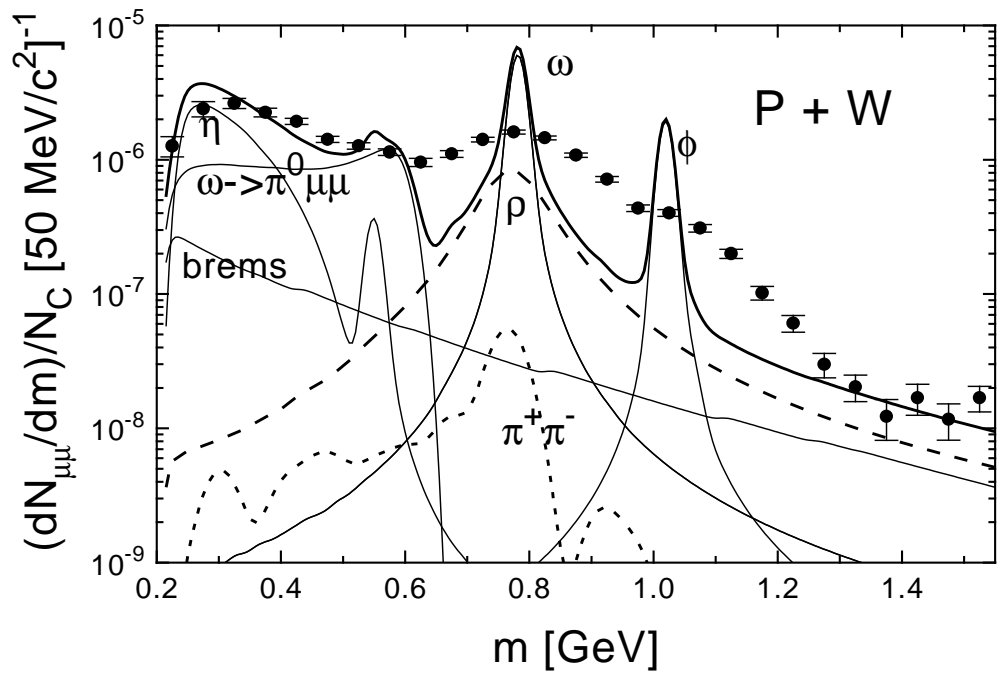
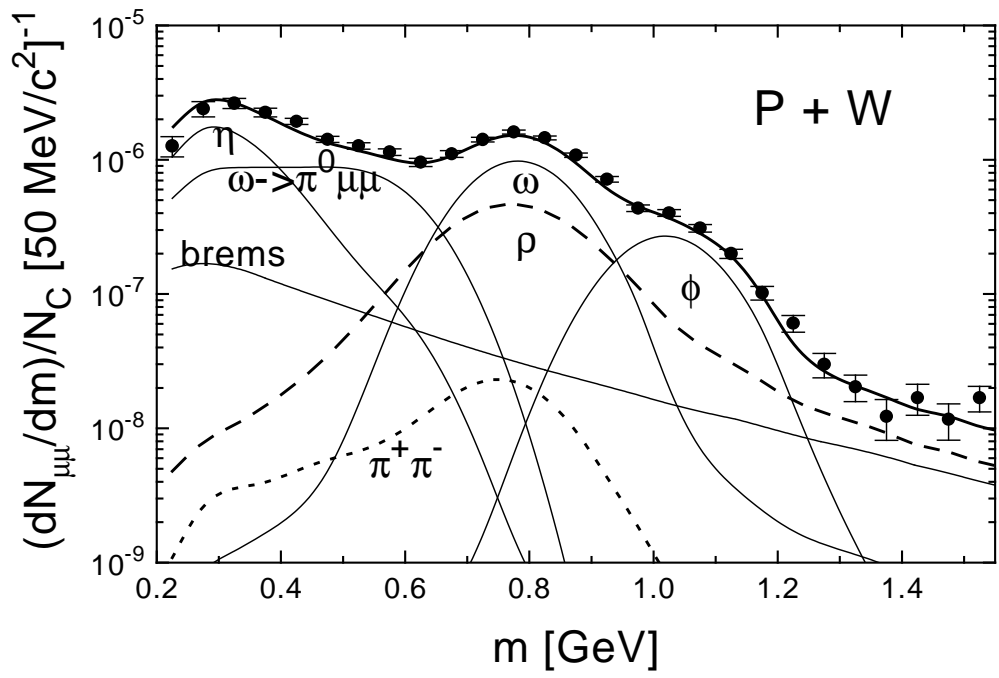
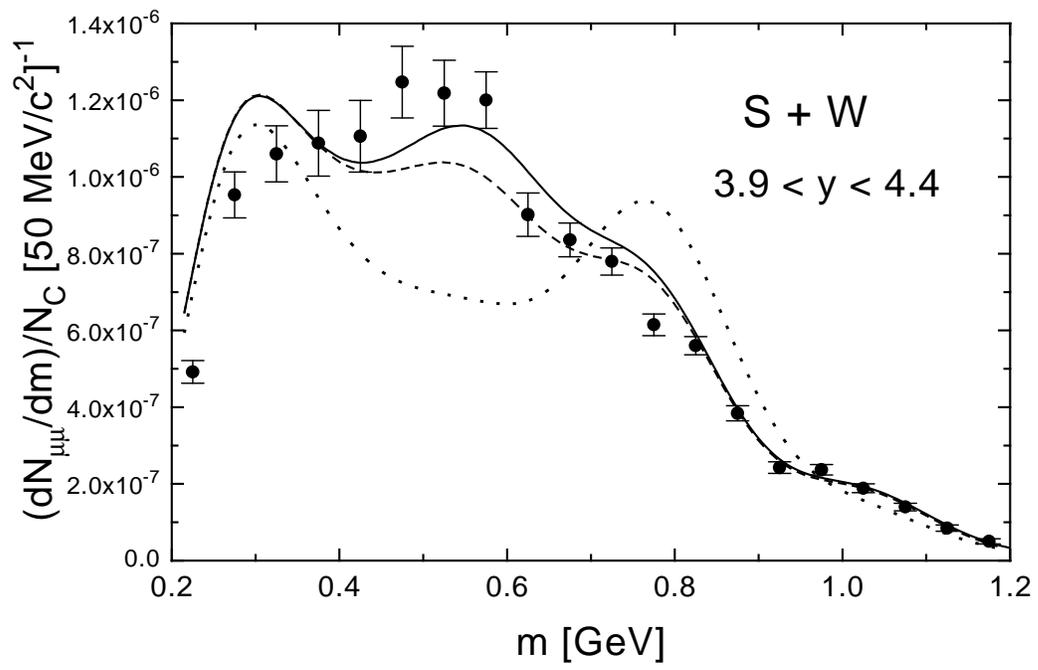
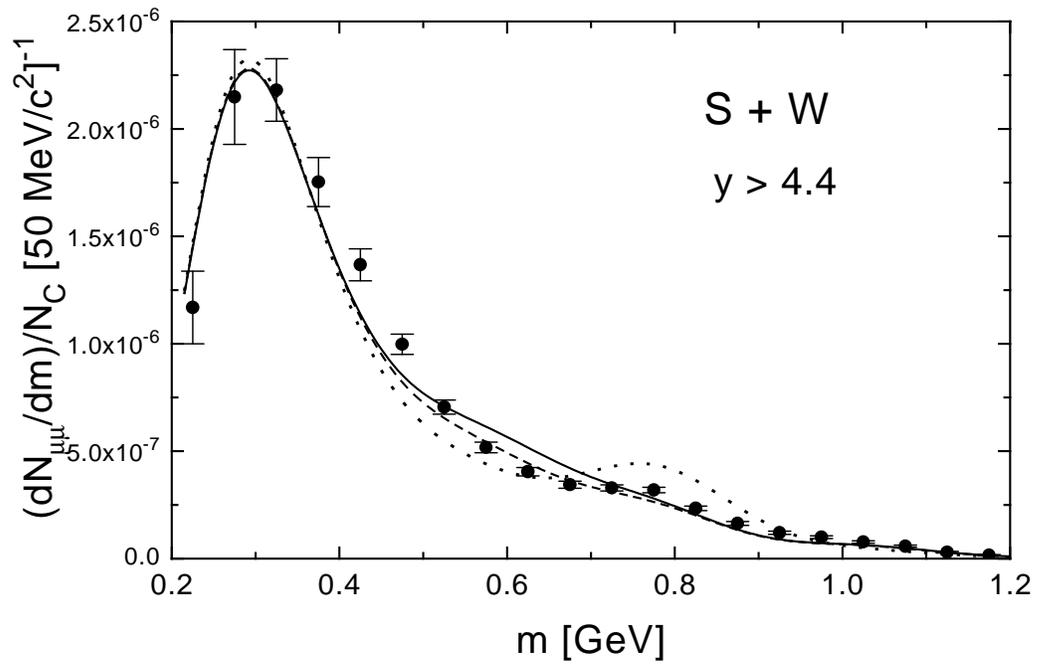
\end{document}